\documentclass[11pt]{article}

\usepackage{cite}
 \usepackage{subfigure}
\usepackage{multirow}
\usepackage{helvet}
\usepackage{amsmath}
\usepackage{amssymb}
\usepackage{setspace}
\usepackage{graphicx}
\usepackage{empheq}
\usepackage{soul}
\usepackage{tabularx}
\usepackage{array}
\usepackage{float}
\usepackage{braket}
\usepackage[utf8]{inputenc}
\usepackage{url}
\usepackage{hyperref}
\usepackage{slashed}
\usepackage[font=footnotesize,labelfont=bf]{caption}
\usepackage{color}

\def\be{\begin{equation}}
\def\ee{\end{equation}}

\usepackage[a4paper,top=3cm,bottom=3cm,left=2.5cm,right=2.5cm,bindingoffset=0mm]{geometry}

\begin{document}
\hspace*{\fill} DESY-25-180 

\begin{center}

{\Large \bf The impact of inclusive electron ion collider data on the strong\\ \vspace*{0.3cm}   coupling determination in a global PDF fit}

\vspace*{1cm}
 L. A. Harland-Lang$^{a}$, T. Cridge$^{b,c}$, P. Newman$^{d}$, R.S. Thorne$^a$ and K. Wichmann$^{e}$ \\                                           
\vspace*{0.5cm}                                                    
$^a$ Department of Physics and Astronomy, University College London, London, WC1E 6BT, UK \\   
$^b$ Elementary Particle Physics, University of Antwerp, Groenenborgerlaan 171, 2020 Antwerp, Belgium \\
$^c$ Department of Physics and Astronomy, University of Manchester, Manchester, M13 9PL, United Kingdom \\
$^d$ School of Physics and Astronomy, University of Birmingham, B15 2TT, United Kingdom \\  
$^e$ Deutsches Elektronen-Synchrotron DESY, Germany

\begin{abstract}
\noindent We present a study of the impact of data from the upcoming Electron Ion Collider (EIC) on the determination of the strong coupling within the context of the global MSHT fitting framework. To achieve this, we generate EIC electron-proton scattering pseudodata according to both conservative and optimistic 
experimental uncertainty projections and perform a simultaneous fit to obtain the proton PDFs and the value of the strong coupling. In the conservative case the impact is found to be moderate, but non--negligible, while in the optimistic case it is observed to be rather significant. These results therefore underline the promising potential for the EIC in the determination of the strong coupling. We in addition explore the impact of any potential tensions between the EIC data and the rest of the data in the global fit by injecting explicit inconsistencies into the pseudodata generation. This can lead to a noticeable bias in the extracted value of the strong coupling, highlighting the importance of accounting for all sources of theoretical uncertainty in the fit as well as the relevance of an enlarged, conservative, error definition in the determination of the strong coupling.

\end{abstract}

\end{center}

\section{Introduction}

Parton distribution functions (PDFs) are an essential part of any theory prediction for
hard hadronic interactions, and are therefore  key to our understanding of data from collider experiments. There has been significant progress in extracting PDFs as accurately and precisely as possible from different groups, either via global PDF fits~\cite{Bailey:2020ooq,NNPDF:2021njg,Hou:2019efy,Alekhin:2017kpj} or more dedicated extractions using data from HERA and the LHC~\cite{H1:2015ubc,ATLAS:2021vod}. Due to the high precision requirements of such fits, next--to--next--to leading order (NNLO) in the QCD perturbative expansion is now the default, but both MSHT  
\cite{McGowan:2022nag} and NNPDF \cite{NNPDF:2024nan} have presented fits at approximate N$^3$LO (aN$^3$LO).

Given the inherently QCD nature of the PDFs, and the interactions underlying the processes that enter these fits, as well as being sensitive to the PDFs, they are also in general sensitive the value of the strong coupling. As such, multiple groups have presented extractions of the value of the strong coupling,  most recently in~\cite{Alekhin:2017kpj,Hou:2019efy,Cridge:2024exf,Ball:2025xgq}. It is of particular note that the quoted uncertainty in these studies is of the same order as, if generally slightly larger than, the PDG average~\cite{ParticleDataGroup:2024cfk}, see also~\cite{dEnterria:2022hzv} for a review.

The Electron-Ion Collider~\cite{Accardi:2012qut} (EIC), 
which is expected to begin science operations at Brookhaven 
National Laboratory in the mid 2030s,
is the next particle accelerator that will be able to explore the 
internal structure of hadrons. 
Intended to operate at high luminosities and with a wide range of beam energy configurations, it promises to revolutionise our understanding of QCD.
The impact of the EIC on proton PDFs has been studied previously 
in~\cite{AbdulKhalek:2021gbh,Khalek:2021ulf,Hobbs:2019gob,Hobbs:2019sut,Arratia:2020azl,AbdulKhalek:2022hcn,Amoroso:2022eow,Armesto:2023hnw,NNPDF:2023tyk}, while the potential for a precise determination of the value of the strong coupling has been studied in~\cite{Cerci:2023uhu}. However, as of yet there has been no analysis 
of the EIC sensitivity to the strong coupling
within the context of a global PDF fit. This arguably provides the most appropriate framework to assess the potential impact of EIC data in the determination of the strong coupling, and in  particular its role in the context of the other constraints that enter a global fit, both on the PDFs and strong coupling.
In this paper we present such a study, specifically in the context of the MSHT global PDF fitting framework, at both NNLO and aN$^3$LO in QCD. We in particular evaluate the expected impact from inclusive DIS data from the EIC, by generating pseudodata with both optimistic 
and conservative  scenarios for the projected experimental uncertainties. 

Studies of this type usually make the implicit assumption of perfect consistency between 
all of the datasets included. However, in practice we typically observe multiple tensions between datasets in global PDF fits~\cite{Watt:2012tq,Hou:2019efy,Kovarik:2019xvh,Bailey:2020ooq,Jing:2023isu,Cridge:2024exf}, which can arise not only due to inconsistencies in newly included data, but also due to theoretical effects or simply issues in previously included data in the global fit as whole. 
With this in mind, tensions may  be expected to exist between the new EIC data when they become available
and the rest of the data 
included in a global fit at that time, without this necessarily reflecting any incompleteness in the experimental determination of the  EIC data. These tensions can moderate or alter the impact of 
the new data in constraining the PDFs and the strong coupling, and so result in important effects that ought to be accounted for.
In this work we therefore also analyse the potential impact of tensions between new EIC data and the rest of the data entering the global fit. We achieve this by considering EIC pseudodata where explicit inconsistencies with respect to the baseline fit are built in. 

In more detail, we consider three different
scenarios. In the first case the pseudodata are generated using aN$^3$LO precision, but the fit is performed at NNLO, which effectively models the impact of missing higher order corrections on the result, and hence acts as a measure of the theoretical uncertainty at NNLO.
In the second case, we perform both the EIC
pseudodata production and the fit at
aN$^3$LO, but generate
the pseudodata with target mass corrections included, whilst excluding the corrections 
from the fit.
In the final case, 
we generate the pseudodata with the NNPDF4.0a${\rm N}^3$LO~\cite{NNPDF:2024nan} set, which exhibits some tension with the result of the MSHT20 fit to the non--EIC data. 
The second and third cases  emulate the possibility of some tension, due to missing theory ingredients or some other unspecified source between the EIC data and the rest of the 
data in the fit. Such tensions 
motivate the use of more conservative, enlarged error definitions, 
similarly to the situation in existing PDF fits, 
for example in the previous determinations of the strong coupling by the MSHT group, most recently in~\cite{Cridge:2024exf}.

The outline of this paper is as follows. In Section~\ref{sec:overview} we present an overview of the pseudodata generation and PDF fitting framework; in Section~\ref{sec:results} we present our results at both NNLO and  aN$^3$LO in subsections~\ref{sec:resnnlo} and~\ref{sec:resan3lo}, respectively, 
while in subsections~\ref{sec:incon_NNLO} and ~\ref{sec:incon} we assess the impact of including explicit inconsistencies between the EIC pseudodata and the baseline PDF fit, at NNLO and  aN$^3$LO, respectively; finally, in Section~\ref{sec:conc}, we conclude.

\section{Fit overview}\label{sec:overview}

The pseudodata are by default generated exactly as in~\cite{Armesto:2023hnw} (see also~\cite{Cerci:2023uhu,Jimenez-Lopez:2024hpj}). In particular, both charged and neutral current $ep$ scattering data are generated, for a range of beam energy configurations. In addition to the statistical uncertainty, the pseudodata are generated with a point--to--point uncorrelated systematic uncertainty of 1.9\%, extending to 2.75\% for $y < 0.01$, and a normalization
uncertainty of 3.4\% is taken. The latter is assumed to be fully
correlated between data with the same $\sqrt{s}$ but is uncorrelated between data with different $\sqrt{s}$. However, similarly to the study of~\cite{Jimenez-Lopez:2024hpj} we also consider an `optimistic' scenario, with rather smaller experimental uncertainties. In particular, for this we take a 0.7\% normalization uncertainty that is fully correlated across bins within beam energies but not across beam energies and a 0.7\% bin--by--bin uncorrelated uncertainty, to be compared with 3.4\% and 1.9\% respectively in the default case.

For the a${\rm N}^3$LO results, we use the same basic framework as in the original MSHT20a${\rm N}^3$LO fit~\cite{McGowan:2022nag}, but we now use the more up--to--date splitting functions~\cite{Falcioni:2023luc,Falcioni:2023vqq,Falcioni:2023tzp,Moch:2023tdj,Falcioni:2024xyt,Falcioni:2024qpd} that have become available following the original MSHT20a${\rm N}^3$LO fit, as described in \cite{Cridge:2025oel}.

Following~\cite{Harland-Lang:2025wvm} we have also updated the treatment of various fixed target datasets. In particular, we now include a more precise account of the correlated systematic uncertainties and separation of the datasets into different beam energies. The updated datasets are the BCDMS~\cite{BCDMS:1989qop,BCDMS:1989ggw}, NMC~\cite{NewMuon:1996fwh,NewMuon:1996uwk} and E665~\cite{E665:1996mob} proton and deuteron data, and NuTeV $F_{2,3}$ structure function data~\cite{NuTeV:2005wsg}. 

\section{Results}\label{sec:results}

\subsection{NNLO QCD Fits}\label{sec:resnnlo}

\begin{figure}
\begin{center}
\includegraphics[scale=0.7]{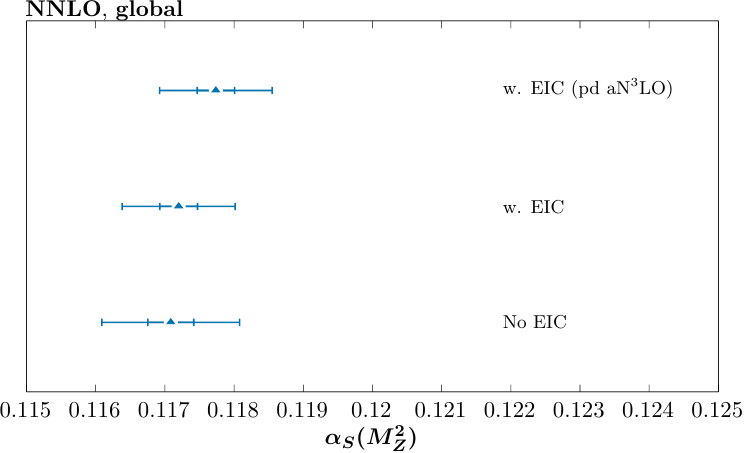}
\includegraphics[scale=0.7]{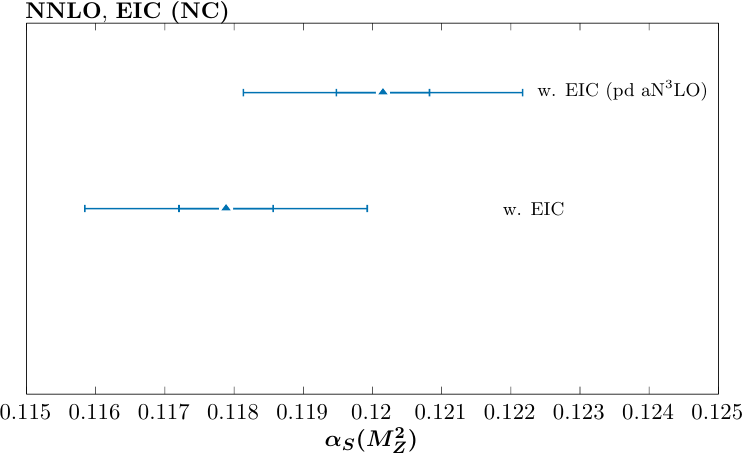}
\caption{\sf Best fit value of the strong coupling $\alpha_S(M_Z^2)$ for a range of NNLO fits, including (`w. EIC') or excluding EIC (`no EIC') pseudodata (`pd'), in the conservative scenario. The inner (outer) limits correspond to $\Delta \chi^2=1 (9)$, and are intended to be indicative of the relative uncertainty on $\alpha_S(M_Z^2)$ in different fits, rather than indicating the actual size of the uncertainty that would result from the dynamic tolerance procedure. The 
top (bottom) plots show the result of the globally and locally (EIC NC) preferred values.
} 
\label{fig:as_NNLO}
\end{center}
\end{figure}

\begin{figure}
\begin{center}
\includegraphics[scale=0.65]{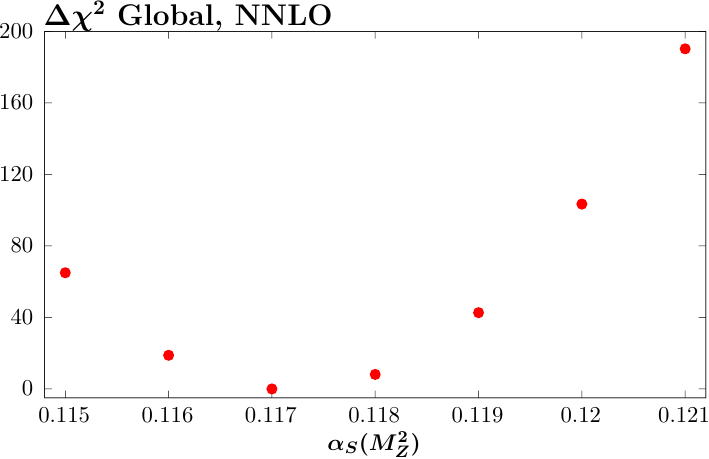}
\includegraphics[scale=0.65]{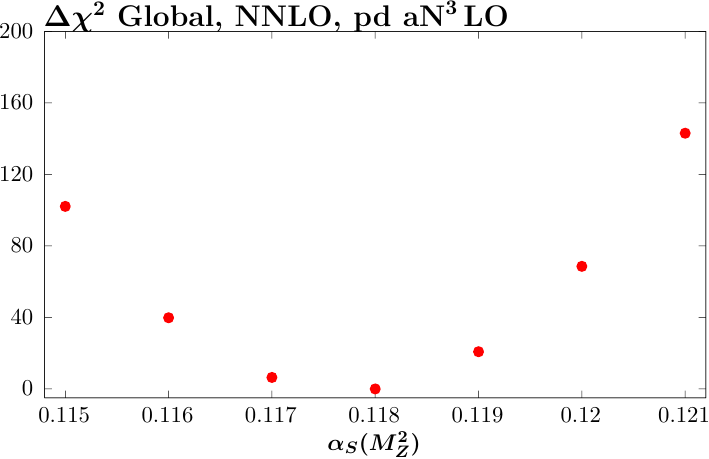}
\includegraphics[scale=0.65]{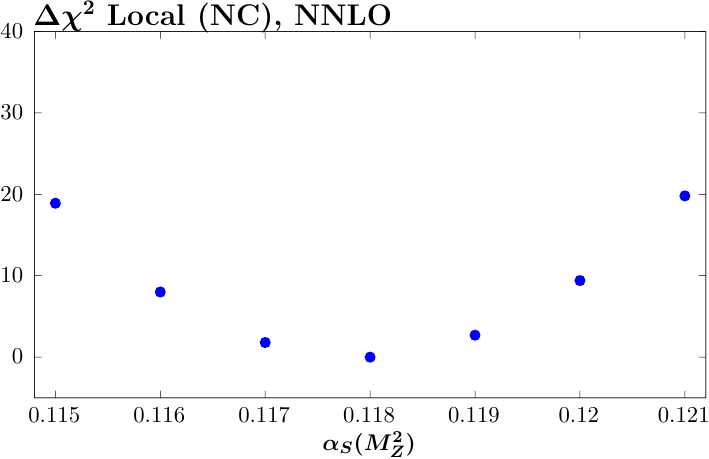}
\includegraphics[scale=0.65]{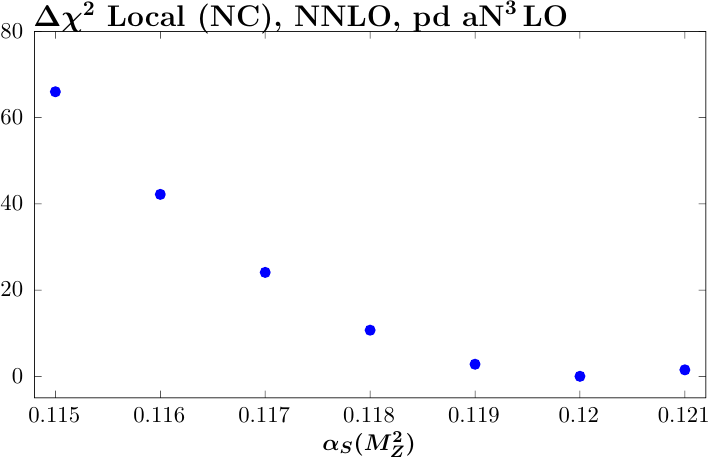}
\caption{\sf $\chi^2$ profiles corresponding to the `w. EIC' and `w. EIC (pd a${\rm N}^3$LO)' cases in Fig.~\ref{fig:as_NNLO}. }
\label{fig:as_NNLO_pdN3LO}
\end{center}
\end{figure}

We begin by considering fits performed at NNLO in the QCD perturbative expansion. The best fit values of the strong coupling, and an estimate of their uncertainties, for a range of fits that we perform are shown in Fig.~\ref{fig:as_NNLO}. The 
top and bottom plots show the result of the globally and locally preferred values, respectively. 
The inner (outer) limits correspond to $\Delta \chi^2=1$ (9), that result from a quadratic fit about the minimum to the corresponding global profile. These are intended to be indicative of the relative uncertainty on $\alpha_S(M_Z^2)$ in different fits, rather than indicating the actual size of the uncertainty that would result from the dynamic tolerance procedure that is used by the MSHT collaboration, as described in~\cite{Martin:2009iq}. To evaluate these, and the corresponding best fit value, a scan over fixed values of $\alpha_S(M_Z^2)$, spaced by 0.001, is performed in each case and the three points closest to the minimum are used for a quadratic fit to the $\Delta \chi^2$ profile. 
When showing local profiles, we only consider the neutral current (NC) data, considering this to have a more direct constraint on the strong coupling, although including the charged current data  in these makes little difference. For illustration, in Fig.~\ref{fig:as_NNLO_pdN3LO}  we show the corresponding $\chi^2$ profiles for the conservative with and without EIC pseudodata. 

We note that the best fit values for the strong coupling in the no EIC case are not the same as in~\cite{Cridge:2024exf}. This is due to the update to the fixed target data and evolution described in Section~\ref{sec:overview}, but also to a corrected treatment of the  a${\rm N}^3$LO  K-factors for the jet and vector boson $p_\perp$ (and + jet) datasets, as discussed further in~\cite{Harland-Lang:2025wvm}. Furthermore, in this study we do not include the SeaQuest data~\cite{SeaQuest:2021zxb} in the fit, and for this reason these results are not identical to~\cite{Harland-Lang:2025wvm}. This overall difference, while noticeable, remains  within the uncertainty due to the dynamic tolerance treatment. Moreover, we emphasise that the aim here is to analyse the impact of the EIC pseudodata with respect to the baseline fit excluding EIC pseudodata, rather than investigate the specific value of the strong coupling.

\begin{table}
\begin{center}
  \scriptsize
  \centering
   \renewcommand{\arraystretch}{1.4}
\begin{tabular}{Xrccc}\hline 
&no EIC & w. EIC & w EIC (pd  a${\rm N}^3$LO)
\\ \hline
{\bf EIC}& 690.0 (1.25) &{\bf 522.5 (0.95)}&{\bf 583.5 (1.06)}\\
{\bf Fixed Target}&{\bf 2168.0 (1.09)} &{\bf 2197.0 (1.11)} &{\bf 2178.0 (1.10)} \\\
{\bf HERA}& {\bf 1623.5 (1.28)}&{\bf 1612.3 (1.28)}&{\bf 1625.6 (1.29)}\\
{\bf Hadron Collider}& {\bf 2513.6 (1.40)}&{\bf 2513.4 (1.40)}&{\bf 2520.0 (1.41)}
\\ \hline \hline 
{\bf Global (no EIC)}  &\bf{6305.2 (1.25)} & \bf{6322.6 (1.26)}  & \bf{6323.7 (1.26)}    \\
{\bf Global}  &6995.1 (1.25) & \bf{6845.0 (1.22)}& \bf{6907.2 (1.24)} \\
\hline
\end{tabular}
\end{center}
\caption{\sf $\chi^2$ values for MSHT fits performed at NNLO, including and excluding 
conservative
EIC pseudodata in different forms, as indicated. The global fit quality as well as that for various subsets is shown. For the fit excluding EIC pd, the predicted $\chi^2$ values are shown (not in bold) for the EIC pd. The $\chi^2$ values are indicated, with the $\chi^2$ per point given in brackets.}
\label{tab:chi2_Fits_NNLO}
\end{table}

Comparing the `No EIC' and `w EIC' cases we see the impact of including the EIC pseudodata when it is  generated consistently at NNLO. Locally the EIC pseudodata prefer $\alpha_S(M_Z^2)\sim 0.118$ as we would expect, given this is used to generate the input. This results in a mild increase in the globally preferred value of the strong coupling, from $\sim 0.1171$ to 0.1172. 

The corresponding fit qualities, i.e. the $\chi^2$ per point,  are shown in Table~\ref{tab:chi2_Fits_NNLO}. Prior to refitting, the description of the EIC pseudodata is, statistically speaking, relatively poor in the comparison. This is permissible, given that while the underlying theory used to generate the pseudodata is broadly consistent with the fit, it is not identical; that is, the underlying MSHT20NNLO set used to generate the pseudodata, as well as the input value of  $\alpha_S(M_Z^2)= 0.118$ are not identical to the result of this, more recent, fit excluding EIC pseudodata. There is in particular no account of PDF uncertainties in the fit qualities shown here. On the other hand, after refitting, we can see that the 
fit quality is indeed close to unity with only a rather mild deterioration in the description of the non--EIC data, and so the pseudodata can be safely accommodated.

\subsection{Impact of explicitly inconsistent EIC pseudodata: NNLO}\label{sec:incon_NNLO}

We next investigate the impact of intentionally injecting some inconsistency between the generated pseudodata and the
existing best fit PDFs arising from the global fit
to existing data. In practice such an inconsistency could
arise due to the EIC data being more sensitive to certain theoretical corrections than the existing data, or due to issues in the experimental or theoretical uncertainty estimations
in older DIS or hadron collider data that 
constrain similar partons to the EIC data.
Introducing the explicit inconsistency thus serves as a proxy for a wide range of possible 
sources of tension in the fits. Ideally such tensions
would be addressed by identifying  
the underlying reason for the inconsistency 
and introducing explicit additional sources of 
uncertainty accordingly. However, where the reasons are not
identifiable, such tensions require caution in the 
interpretation of the fit results.

In Fig.~\ref{fig:as_NNLO} we show the `w. EIC (pd a${\rm N}^3$LO)' result where the pseudodata are instead generated at a${\rm N}^3$LO, while maintaining a NNLO fit; in other words, such that the pseudodata and the fit itself are by construction inconsistent. The pseudodata now locally prefer a higher value of $\alpha_S(M_Z^2)\sim 0.120$, despite the fact that these are generated at  $\alpha_S(M_Z^2)= 0.118$. That is, the inconsistency between the a${\rm N}^3$LO generated pseudodata and the NNLO fit results in a bias in the preferred value of the strong coupling for the EIC pseudodata, with the larger value effectively compensating for the missing higher order between the NNLO and a${\rm N}^3$LO description. Globally, this is seen to  pull the preferred value of  $\alpha_S(M_Z^2)$ up from $\sim 0.1171$ to $\sim 0.1176$. These differences are on the edge of the textbook $\Delta \chi^2=T^2=1$ uncertainty, but well within the enlarged $T=3$ tolerance more typical of the MSHT dynamic tolerance treatment. At the level of the fit quality, however, we can see from Table~\ref{tab:chi2_Fits_NNLO}
that the description of the EIC pseudodata remains completely acceptable\footnote{A $1\sigma$ variation in the $\chi^2/N$ for the number of points is $\sqrt{2/N}\sim 0.06$, and hence this is only $\sim 1\sigma$ above unity.}. This therefore indicates the need for caution when assessing the level of incompatibility between datasets from the overall fit quality alone. In particular, while this is not at all evident in the $\chi^2$ deterioration, the impact on the extracted value of the strong coupling, if a naive $T^2=1$ approach is taken, is not negligible. We discuss this point further in Section~\ref{sec:incon}, in the context of the a${\rm N}^3$LO fits.

\subsection{a${\rm N}^3$LO QCD Fits}\label{sec:resan3lo}

\begin{figure}[t]
\begin{center}
\includegraphics[scale=0.8]{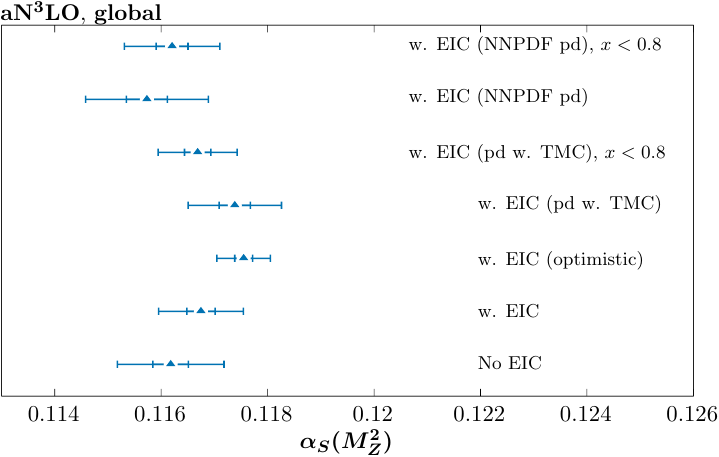}
\includegraphics[scale=0.8]{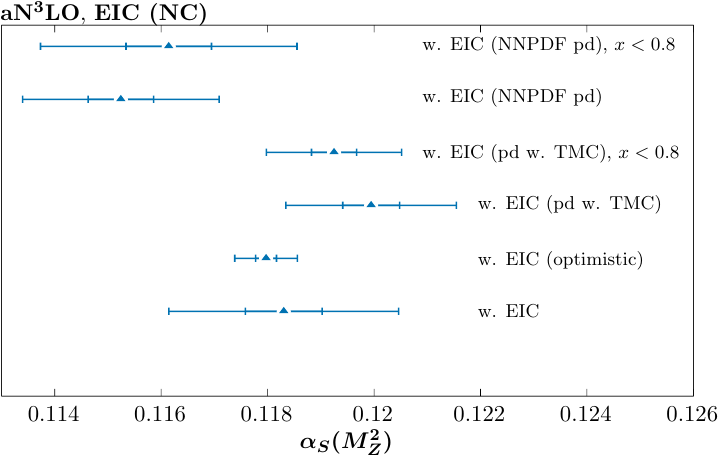}
\caption{\sf Best fit value of the strong coupling $\alpha_S(M_Z^2)$ for a range of a${\rm N}^3$LO  fits, including or excluding EIC pseudodata. The inner (outer) limits correspond to $\Delta \chi^2=1$ (9)  that result from a quadratic fit about the minimum to the corresponding global profile, and are intended to be indicative of the relative uncertainty on $\alpha_S(M_Z^2)$ in different fits, rather than indicating the actual size of the uncertainty that would result from the dynamic tolerance procedure. The left (right) plots show the result of the globally and locally (EIC NC) preferred values. Unless otherwise specified, the pseudodata correspond to the conservative scenario}
\label{fig:as_N3LO}
\end{center}
\end{figure}

\begin{figure}
\begin{center}
\includegraphics[scale=0.65]{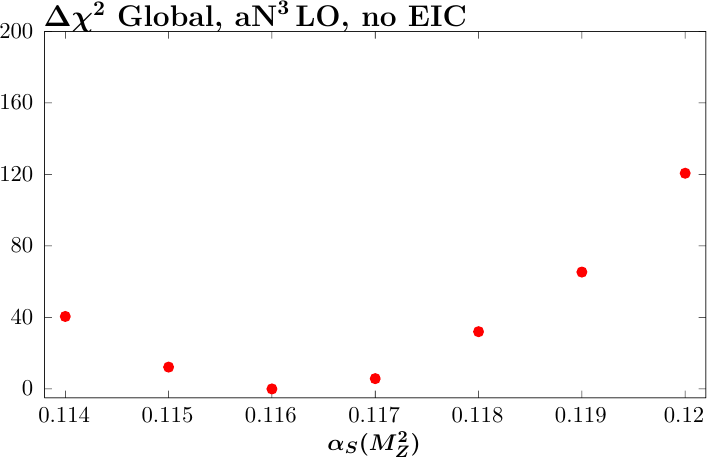}
\includegraphics[scale=0.65]{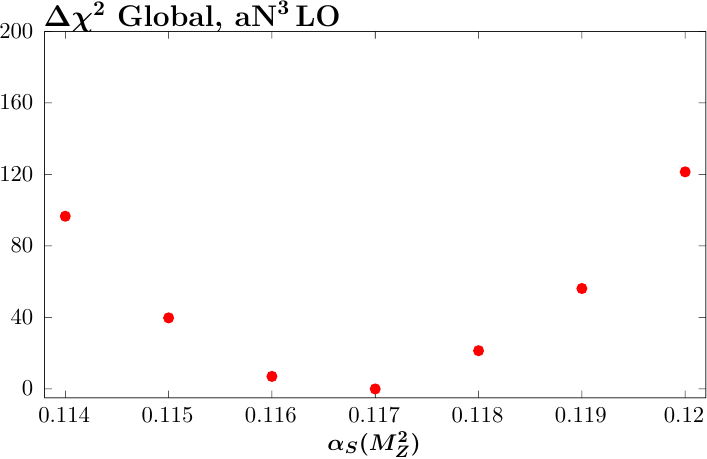}
\includegraphics[scale=0.65]{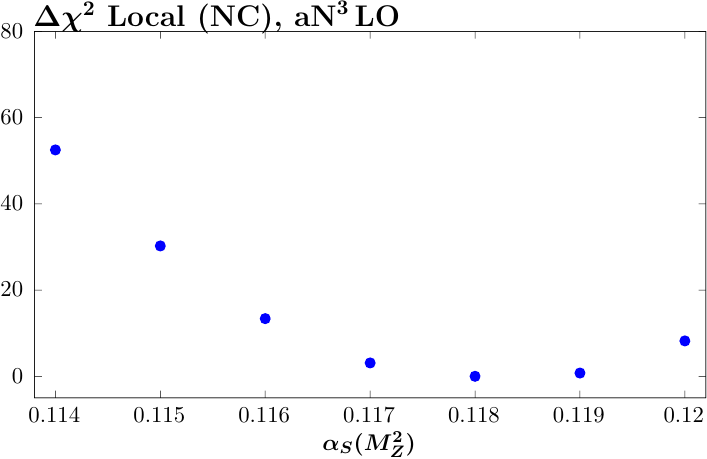}
\caption{\sf $\chi^2$ profiles corresponding to the `No EIC' and `w. EIC' cases in Fig.~\ref{fig:as_N3LO}.}
\label{fig:as_N3LO_pdN3LO}
\end{center}
\end{figure}

In this section we consider the a${\rm N}^3$LO fit. The fit quality is shown in Table~\ref{tab:chi2_Fits}.
Before refitting, the description of the EIC pseudodata is good, being less than $ 2\sigma$ above unity.  
While an acceptable level of deviation due to purely statistical scatter alone, we note that as discussed above in the case of the NNLO fit, while the underlying PDFs used to generate the pseudodata are broadly consistent with those extracted from the baseline fit, they are not identical, and hence some deviation may be expected.
After refitting, on the other hand, we can see that the $\chi^2$ per point is close to unity with only a rather mild deterioration in the description of the non--EIC data, and so the pseudodata can again be safely accommodated.

In terms of the impact on the strong coupling determination, shown in Fig.~\ref{fig:as_N3LO}, the results for the conservative scenario are rather similar to the NNLO case. We  show the corresponding $\chi^2$ profiles with and without EIC pseudodata (conservative scenario) in Fig.~\ref{fig:as_N3LO_pdN3LO}. The locally preferred value of the EIC pseudodata remains at 0.118, and this increases the globally preferred value of the strong coupling somewhat. In more detail, we observe that the global impact of the EIC pseudodata on the best fit value of the strong coupling is more significant
than in the NNLO case, increasing it from $\sim 0.1162$ to $\sim 0.1167$. 

The impact on the uncertainty of the strong coupling
can be evaluated following the standard dynamic tolerance criterion discussed in e.g.~\cite{Bailey:2020ooq}. As the locally preferred value is rather higher than the global best fit value, we expect the lower limit to be more stringent. This is indeed the case, with a lower limit corresponding to a shift
of $\sim -0.0015$, i.e. comparable to existing lower limits in the most recent MSHT study~\cite{Cridge:2024exf}, which come from lower energy fixed--target data. We do not perform a full reanalysis here, as the precise uncertainty on the coupling will depend on factors such as how the pseudodata are generated; if the input value of the coupling is more in line with the globally preferred one, for example, this will change the precise limits. Nonetheless, we broadly find that the expected limit is of the order of existing ones. Given the potentially better understanding of the detector and more refined treatment of experimental uncertainties that we may expect at the EIC, this is encouraging. 
The pull on the global value of the coupling is rather larger in comparison to the NNLO case since the preferred value at a${\rm N}^3$LO is somewhat lower than at NNLO, and lower than in the recent study~\cite{Cridge:2024exf}, as discussed above.  That is, the central value for the baseline without EIC pseudodata is rather further away than the value of 0.118 that we have used to generate the pseudodata than at NNLO, leading to a larger expected shift.

\begin{table}
\begin{center}
  \scriptsize
  \centering
   \renewcommand{\arraystretch}{1.4}
\begin{tabular}{Xrcccc}\hline 
&no EIC & w. EIC & w EIC (TMC) &w EIC (NNPDF4.0) 
\\ \hline
{\bf EIC}& 572.3 (1.10) &{\bf 509.2 (0.98)}&{\bf 532.0 (1.02)}&{\bf 584.9 (1.12)}\\
{\bf Fixed Target}&{\bf 2153.6 (1.09)} &{\bf 2155.3 (1.09)} &{\bf 2149.8 (1.08)} &{\bf 2215.9 (1.29)}\\\
{\bf HERA}& {\bf 1626.3 (1.29)}&{\bf 1629.3 (1.29)}&{\bf 1631.1 (1.29)}&{\bf 1633.9 (1.29)}\\
{\bf Hadron Collider}& {\bf 2388.1 (1.33)}&{\bf 2390.3 (1.34)}&{\bf 2388.7 (1.33)}&{\bf 2432.4 (1.36)}
\\ \hline \hline 
{\bf Global (no EIC)}  &\bf{6168.0 (1.23)} & \bf{6175.0 (1.23)}  & \bf{6169.6 (1.23)} & \bf{6282.2 (1.25)}   \\
{\bf Global}  &6740.3 (1.21) & \bf{6684.2 (1.20)}& \bf{6701.6 (1.21)}& \bf{6787.0 (1.24)}  \\
\hline
\end{tabular}
\end{center}
\caption{\sf $\chi^2$ values for MSHT fits performed at a${\rm N}^3$LO, including and excluding EIC pseudodata in different forms, as indicated. The global fit quality as well as that for various subsets is shown. For the fit excluding EIC pd, the predicted $\chi^2$ values are shown (not in bold) for the EIC pd. The `TMC' and `NNPDF4.0' cases correspond to generating the pseudodata with explicit inconsistencies, as described in the text.}
\label{tab:chi2_Fits}
\end{table}

We next consider the impact of the EIC pseudodata generated in the optimistic scenario described in the  Section~\ref{sec:overview}. The difference in the local profile is significant, decreasing the fixed $T$ uncertainty by a factor of
$\sim 3-4$. 
Globally, this is washed out to some extent, as we would expect, but nonetheless the fixed tolerance uncertainty is reduced by $\sim 50\%$. This therefore demonstrates the clear potential for a large impact on the strong coupling determination by EIC data. Applying the dynamic tolerance procedure, we find that the lower limit on the strong coupling is greatly reduced with respect to previous MSHT results, being a shift of $\sim -0.005$ (i.e. rather in line with the fixed $T^2=9$ result shown in the figure), while the upper limit is  {a shift of $\sim 0.0015$. As always the precise values will depend on the details of the real data, but this again indicates the strong potential constraining power of such data.

\subsection{Impact of explicitly inconsistent EIC pseudodata:  a${\rm N}^3$LO}\label{sec:incon}

We next, as in Section~\ref{sec:incon_NNLO} but now at a${\rm N}^3$LO, investigate the impact of intentionally injecting some inconsistency between the generated pseudodata and the
existing best fit PDFs arising from the global fit
to existing data. 

First, the `EIC (pd w. TMC)' result shows the impact of generating the pseudodata with theoretical predictions that include target mass corrections (TMCs) in $F_2$, as outlined in e.g.~\cite{Schienbein:2007gr}, see Eq.~(61) of this reference, and considered recently in~\cite{Harland-Lang:2025wvm} . For the fit, however, we omit these; while in principle these can (and should) be included, where relevant this can provide some handle on any missing theoretical ingredients in the calculation of this sort. Interestingly, this is seen to result in a significant upwards shift in the locally preferred value of the strong coupling, and due to this a further increase in the globally preferred value, from $\sim 0.1162$ to $\sim 0.1174$ in comparison to the result without EIC pseudodata, and from $\sim 0.1167$ to $\sim 0.1174$ in comparison to the result with EIC pseudodata in the consistent case. This is despite the fact that the overall fit quality, shown in Table~\ref{tab:chi2_Fits}, does not noticeably deteriorate. This is not a contradiction, given the difference between the parameter fitting criterion one in principle uses to evaluate the best fit value of the strong coupling and its uncertainty, in comparison to the hypothesis testing criterion used to evaluate the overall fit quality. Nonetheless, this clearly indicates the danger of naively assuming an e.g. $\Delta \chi^2=1$ criterion can be used, based on observation of the overall fit quality being reasonable.
 In particular, the global $\Delta \chi^2=1$ uncertainty in the current case is $\sim \pm 0.0003$, and hence the above shift would naively correspond to a $4\sigma$ shift. For the arguably more relevant comparison to the preferred value when the EIC pseudodata are consistently included, for which the preferred value of the strong coupling is $\sim 0.1167$, this corresponds to a $\sim 2\sigma$ shift with respect to this.

However, the reason for this shift lies almost entirely in the $x = 0.815$ bins that exist in the pseudodata, which are at sufficiently high $x$ that even for the reasonably high values of $Q^2$  ($\sim 100\,{\rm GeV}^2$ or more) probed, TMCs are relevant. Indeed, by simply removing these bins we end up with a result that is again consistent with the result from the fit with consistent pseudodata, i.e. the biasing effect is largely removed in the global fit (although it remains somewhat visible in the local profile).

To investigate another possible
source of inconsistency, we instead generate the pseudodata with the NNPDF4.0a${\rm N}^3$LO ~\cite{NNPDF:2024nan} set. As shown in Fig.~\ref{fig:msht_vs_nnpdf_chw} for the relevant case of the charge weighted quark singlet, there are some regions of $x$ where these are in tension, even beyond the quoted uncertainty bands. This therefore provides a method to model the impact of a situation where the eventual EIC data prefer a somewhat different shape to the underlying PDFs, while being agnostic as to the underlying reason. 

The results are shown in Fig.~\ref{fig:as_N3LO} 
as the `w. EIC (NNPDF pd)' cases. The local profile is now shifted towards lower values of the strong coupling, albeit with a relatively shallow minimum. The reason for this is most likely due to the fact that for the smaller NNPDF4.0 charge weighted quark singlet, the corresponding EIC pseudodata are in general lower and hence a smaller coupling is required to reduce the theoretical prediction and match these pseudodata. This leads to a shift downwards in the globally preferred  value of the strong coupling, from $\sim 0.1162$, the value without  EIC pseudodata included, to $\sim 0.1157$. This shift is $\sim 1- 2\sigma$ if a textbook $\Delta \chi^2=1$ uncertainty is applied, corresponding to a $\sim 0.0003$ variation in the strong coupling. In fact, we recall from above that if the pseudodata are applied without any explicit inconsistency then the preferred value of the strong coupling is shifted upward to $\sim 0.1167$ (see Section~\ref{sec:resan3lo}), i.e. an increase by $\sim 1-2\sigma$ towards the value of 0.118 used to generate the pseudodata. In other words, this change in the preferred value of the strong coupling is of the same order of shift as the case with  consistently generated pseudodata. 
However, due to the tension now present, this shift is in the opposite direction to the consistent case, that is {\it away} from the value of the strong coupling of 0.118 that is used to generate the pseudodata.

\begin{figure}
\begin{center}
\includegraphics[scale=0.65]{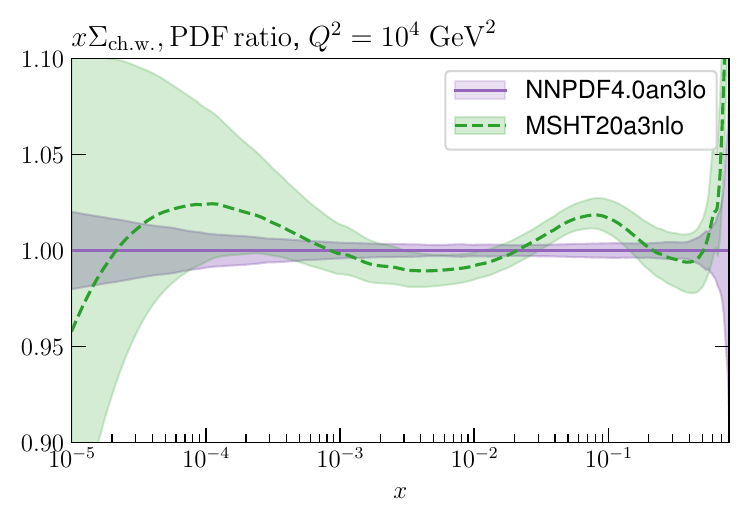}
\caption{\sf Ratio of the MSHT20a${\rm N}^3$LO to NNPDF4.0a${\rm N}^3$LO charge weighted quark singlet PDFs at $Q^2=10^4\,{\rm GeV}^2$.}
\label{fig:msht_vs_nnpdf_chw}
\end{center}
\end{figure}

These results indicate  the manner in which tensions between datasets can lead to differences in the preferred value of the strong coupling. We   note that  the overall fit quality, shown in Table~\ref{tab:chi2_Fits}, deteriorates by $\sim 2\sigma$, which is therefore noticeable but not unusual in the context of global PDF fits. If the dynamic tolerance criterion is applied we find that, due in part to the flatter local profile but also the precise position of the preferred value of the coupling, the constraints due to the EIC pseudodata are looser than in the consistent case, with the (smaller) upper limit being $\sim +0.002$.

If we remove the $x=0.815$ data points, as above, then we may expect from Fig.~\ref{fig:msht_vs_nnpdf_chw} that this will moderate the tension between the EIC pseudodata generated with NNPDF4.0 and the rest of the global fit, given the largest difference between the NNPDF4.0 and the MSHT20 set for the charge weighted quark singlet is in this very high $x$ region. The result of doing this is shown in Fig.~\ref{fig:as_N3LO}, and indeed there is a tendency to reduce the shift towards lower values of the strong coupling in the local profile; consequently, globally the central value of the strong coupling is $\sim 0.1162$, i.e. essentially unchanged with respect to the case without EIC pseudodata. 

However, both of these cases should also be considered in light of the result for the optimistic scenario. We can see in Fig.~\ref{fig:as_N3LO} that the result of generating the pseudodata with the NNPDF4.0 set is in significant tension with the EIC (optimistic) result, when using the $T^2=1$ uncertainty. This is however rather reduced (though still somewhat present) if an enlarged tolerance is used. For the TMC case, the upwards shift in fact accidentally moves the result close to the optimistic case. However the size of the shift from this effect is noticeable with respect to the $T^2=1$ uncertainty. Clearly if the inconsistent pseudodata exercises were performed within the optimistic scenario, the level of tension would increase further.

\section{Summary and Outlook}\label{sec:conc}

In this paper we have presented a study of the impact that EIC measurements of inclusive DIS have on the determination of the strong coupling, within the context of the MSHT global PDF fitting framework, at both NNLO and aN$^3$LO in QCD. We have supplemented the existing MSHT global dataset with EIC pseudodata and performed a simultaneous PDF and $\alpha_S$ determination. We have performed this study with both a conservative and more optimistic set of assumptions about the eventual experimental uncertainties on the EIC data. We have found their impact 
 in the conservative case
to be  moderate, but non--negligible, while in the optimistic scenario EIC data lead to a significant reduction in the projected uncertainty. 

We have in addition extended the standard approach to pseudodata projection studies to allow for the possibility of some tension between the EIC data and other data in the fit, which could
occur due to a wide range of theoretical or experimental 
effects that are not fully treated in the uncertainty
determinations, even under the assumption that the EIC
data are perfectly correct within their uncertainties.
This is arguably more representative of the genuine situation in a global PDF fit, where such tensions often occur. To this end, we have considered a range of approaches to inject a source of tension into the pseudodata, namely: by generating the pseudodata at aN$^3$LO order, but fitting at NNLO; by including target mass corrections in the pseudodata but not in the fit; and by generating the pseudodata with the  NNPDF4.0a${\rm N}^3$LO set, which exhibits some difference relative to MSHT. 

We have found that for the inconsistent fits considered above, the overall fit quality to the EIC pseudodata can be perfectly acceptable, but even in the conservative uncertainty scenario lead to shifts of $\sim 1-2\sigma$ in the value of the strong coupling, if a 
textbook $\Delta \chi^2=1$ criterion is used. In the optimistic scenario the relative impact would be larger still. This provides further evidence in support of the need of a complete account of theoretical uncertainties, and of a more conservative approach to error determination, as is advocated by MSHT, in the context of a fit where departures from textbook statistics are evident.

More broadly, however, our results  underline the promising and exciting potential for the EIC in the determination of the strong coupling, and QCD analyses more generally. This is particularly evident under more optimistic assumptions about the eventual experimental uncertainties, subject to all other sources of theoretical uncertainty being under control.

\section*{Acknowledgements}

L. H.-L. and R. S. T. thank the Science and Technology Facilities
Council (STFC) for support via grant awards ST/T000856/1 and ST/X000516/1. T.C. acknowledges that this work has been supported by
funding from Research Foundation-Flanders (FWO) (application number: 12E1323N) and by the Royal Society through Grant URF$\backslash$R1$\backslash$251540.

\bibliography{references}{}

\bibliographystyle{h-physrev}

\end{document}